\documentclass[pre,letterpaper]{revtex4}
\usepackage{amsbsy}
\usepackage{amssymb}
\usepackage{subfigure}
\usepackage{graphicx}

\begin{document}

% \markboth{P.~\v{S}ulc, A.~Wagner, O.C.~Martin}
% {Slow evolutionary dynamics of RNA structures}

%\catchline{}{}{}{}{}
%
%%%%%%%%%%%%%%%%%%%%%%%%%%%%%%%%%%%%%%%%%%%%%%%%%%%%%%%%%%%%%%%%%%%%

\title{QUANTIFYING SLOW EVOLUTIONARY DYNAMICS IN RNA FITNESS LANDSCAPES}

\author{P. \v{S}ULC}
\affiliation{Univ Paris-Sud, UMR8626, LPTMS, Orsay, F-91405; 
\\CNRS, Orsay, F-91405, France\\ petr.sulc@polytechnique.edu}

\author{A.~WAGNER} 
\affiliation{University of Z\"{u}rich, Department of Biochemistry \\
Winterthurerstrasse 190 \\
CH-8057  Z\"{u}rich, Switzerland \\
aw@bioc.uzh.ch
\\
\\
The Santa Fe Institute \\ 1399 Hyde Park Road,
Santa Fe, NM 87501, USA
}

\author{O.~C.~MARTIN}

\affiliation{Univ Paris-Sud, UMR8626, LPTMS, Orsay, F-91405; 
\\CNRS, Orsay, F-91405, France\\ olivier.martin@u-psud.fr}

%\begin{history}
%\received{(Day Month Year)}
%\revised{(Day Month Year)}
%\accepted{(Day Month Year)}
%\comby{(xxxxxxxxxx)}
%\end{history}

%opening

% 
% \topmargin 0mm
% \oddsidemargin 5mm
% \evensidemargin 5mm
% \textwidth 150mm
% \textheight 222mm
% \parindent 5mm
% \marginparwidth 0mm
% \marginparsep 0mm
% \marginparpush 0pt
% \columnwidth \textwidth

\begin{abstract}
We re-examine the evolutionary dynamics of RNA secondary structures under 
directional selection towards an optimum RNA structure. We find that the
punctuated equilibria lead to a very slow approach to the optimum, following
on average an inverse power of the evolutionary time. 
In addition, our study of the trajectories 
shows that the out-of-equilibrium effects due to the 
evolutionary process are very weak. In particular, 
the distribution of genotypes is close to that
arising during equilibrium stabilizing selection. As a consequence,
the evolutionary dynamics leave almost no measurable out-of-equilibrium
trace, only the transition genotypes (close to the
border between different periods of stasis) have atypical mutational properties.
\end{abstract}

\maketitle

\section{Introduction}

In many realistic systems, dynamics undergo a pronounced slow-down, a feature
characteristic of ``complex'' landscapes~\cite{gavrilets}.
Such phenomena arise for instance in physical systems (thermal relaxation~\cite{young}), 
in optimisation problems (diminishing returns 
on search efforts~\cite{hoosStutzle}),
and in evolutionary dynamics (punctuated equilibria,
i.e., long periods of stasis in 
the evolutionary record~\cite{eldredge}). In this work, inspired by the 
theoretical framework of statistical physics for glassy systems, we reconsider a
simple model of evolutionary dynamics, namely RNA secondary structure
evolution~\cite{huynenStadler1996,fontanaSchuster1998a,fontanaSchuster1998b}. 
We find: (i) slow evolutionary dynamics, whereby the time to
find an advantageous phenotypic change
has a distribution with a fat (power-law) tail; 
(ii) non-self-averaging
behavior, \emph{i.e.}, even for long RNA molecules, directional selection for
some targets will lead to significantly slower dynamics than for other targets;
(iii) only weak out-of-equilibrium effects: the genotypes
visited by an evolutionary trajectory are quite similar to those
arising in equilibrium under stabilizing selection, so 
except near the ends of periods of stasis,
the genotypes produced during an evolutionary trajectory do not have 
anomalously high or low mutational robustness.

The paper is organised as follows. In Sect.~\ref{sect:rnaEvolution}
we introduce the evolutionary 
model. In Sect.~\ref{sect:slow} we exhibit
the non-exponential nature of the relaxation process; empirically,
relaxation seems to follow an inverse power law.
We also show that the relaxation curves remain sensitive to the target
used for directional selection even in the limit of very long molecules.
Finally in Sect.~\ref{sect:innovation} 
we give evidence that at nearly all times the evolutionary trajectory is in
quasi-equilibrium; more precisely, the ``innovative'' genotypes produced by
a transition to a new period of stasis
are only a bit different from random genotypes as measured by the
phenotypic effects of mutations; furthermore, during the periods of stasis,
the initial genotype seems to be quickly forgotten, so 
no significant trace of innovation seems to be maintained.
Concluding remarks are given in the
final section.

\section{The RNA evolutionary model}
\label{sect:rnaEvolution}

\subsection{RNA structure and its fitness landscape}
\label{subsect:rnaGenoPheno}

\begin{figure}[t]
 \begin{center}
 \includegraphics[width=6cm, height=8cm, angle=0]{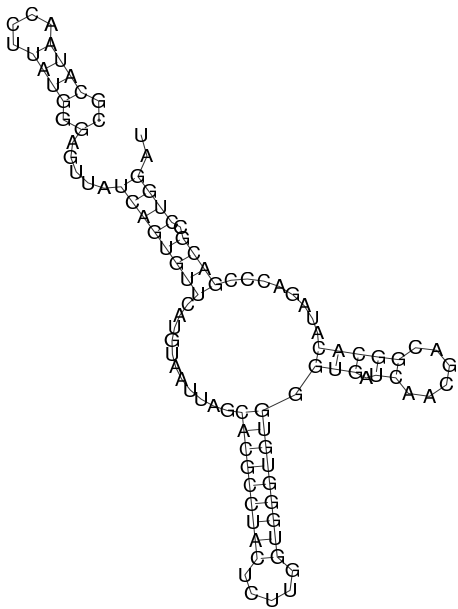}
  \end{center}
 \caption{Example of a secondary structure of an RNA molecule. Several
typical features characteristic of RNA secondary 
structures are shown. They include hairpin loops at the periphery and a multiloop at the center.
The structure shown is represented by
(.((((....)))))......((((((((...........(((((((((.....))))))))).(((.((......)))))........)))).))))..
in dot-bracket notation.
\label{fig:structure}
}
\end{figure}

For the purpose of this study, RNA molecules will be thought of as chains or strings of $L$
nucleotides, taken from an alphabet of four possible nucleotides (A, C, G
and U). Chemically, the bases can pair via hydrogen bonds.
In addition to Watson-Crick pairings (A-U and G-C), the U-G pairing is also
possible, though it is weaker. The pairings
between bases give rise to an RNA secondary structure, as 
illustrated in Fig.\ref{fig:structure}. Apart from this graphical representation,
the secondary structure of an RNA molecule can be specified
by the more convenient dot-bracket notation where a non-paired
base of the sequence is denoted by a dot ``.'' and a paired base is denoted
by a left or right parenthesis. This representation allows one to reconstruct the
pairings as long as the secondary structure is
``planar''~\cite{gesteland}. 
(Planarity means that if the $L$ bases are positioned on a line and
the pairings are represented by arcs between the bases, these arcs can be drawn 
in the plane without any crossings.) An RNA molecule will
spontaneously (\emph{i.e.}, via thermodynamic forces only)
fold into the structures
with lowest free energies. To simulate this folding 
in silico, we use the ``Vienna RNA package''~\cite{vienna}
to find, for any given sequence, the pairings (secondary structure)
which lead to the \emph{minimum} free energy. In effect, this procedure produces a
map from sequences (genotypes) to secondary structures 
(phenotypes)~\cite{fontanaStadler1993,huynenStadler1996,fontanaSchuster1998a,fontanaSchuster1998b}. 
The CPU time needed to determine a minimum free energy structure is $O(L^3)$ for a
chain of $L$ bases.

All genotypes that have the same phenotype form 
a \textit{neutral network}~\cite{fontanaStadler1993}. 
More precisely, given a genotype to phenotype map,
the neutral network associated with the phenotype $P$ 
is a graph whose nodes are all the
genotypes having that phenotype; the edges of that graph
connect genotypes only if they are nearest neighbors.
It is thus necessary to introduce a notion of neighborhood
in genotype space. By convention, two RNA genotypes (sequences)
will be considered as nearest neighbors if and only 
if they differ by a single base.
It is also possible to define a distance $d_p$ between two \emph{phenotypes}. In our case, 
we take the distance between two RNA secondary structures $P_1$ and $P_2$ to be 
$1/L$ times the Hamming distance between their dot-bracket string representations. From this, one 
can introduce a fitness
landscape \cite{stadler}, where genotype and phenotype spaces both have a distance
metric, and where each phenotype can be assigned ``fitness''. We take
the fitness of a phenotype to be a monotonically decreasing function of
its distance to some ``target'' phenotype $P_t$. For our purpose, we 
will extend the notion of a neutral network to 
that of a \textit{neutral ensemble}: 
we call neutral ensemble the union of all 
neutral networks whose corresponding phenotype
has a given fitness, regardless of the specific phenotype. This definition 
reflects that fact that secondary structures are relevant in the 
evolutionary search mainly via their distance to $P_t$.
RNA sequences with their associated fitness form a fitness landscape with
hills, valleys and passes (saddles). Analogous landscapes also 
arise in other systems; for example spin
glasses~\cite{monthusBouchaud96}
are physical materials in which energy is often identified with fitness;
the landscape's ruggedness and many valley structure are important for the
associated dynamical properties.

\subsection{Dynamics under directional selection}
\label{subsect:rnaEvolDyn}

We take the unit of time to be the expected waiting 
time $\Delta t$ between two mutations. Thus if $\mu$ is the mutation rate per RNA molecule
and per generation, our unit of time is $\Delta t = 1/\mu $. For our evolutionary dynamics, 
we simulate the process of RNA evolution towards a target structure by 
allowing a genotype to change by 
a point mutation at each unit of time. 
The directional selection associated with this
evolutionary search then proceeds as follows.
At each step, the genotype is mutated at one base taken at random;
the corresponding phenotype (secondary structure for that genotype)
is determined, and selection is applied: if the distance to the target
structure has increased, the mutation is refused and the previous
genotype is reinstated, otherwise the new genotype is accepted.
This process is usually referred to as ``blind ant'' dynamics~\cite{hughes}
or as an adaptive walk~\cite{levin}. 
(If one were to keep track only
of the accepted moves that are neutral, \emph{i.e.}, that do 
not change the distance to the target phenotype, one would have a random
``neutral'' walk performing myopic ant dynamics~\cite{nimwegen}.) 
In practice, we continue
the random walk until the target structure is reached, or until a maximum number of 
trial mutations is reached.  

In our work, we allow only neutral or improving moves in the landscape, which corresponds to zero
temperature dynamics in physical systems, or hill climbing in optimisation
theory. The trajectories
are stochastic and are influenced both by the initial genotype
chosen (e.g. an arbitrary RNA \emph{sequence}), and by the target (an 
arbitrary \emph{secondary structure}). We thus need to 
average over many trajectories, and consider the dependence of our results on these choices.

In a more general context, one can consider the evolution of a \emph{population} in 
the fitness landscape. If
$\mu$ is the mutation rate and $N$ the population size, then the
effective number of genotypes in a population scales approximately linearly 
with $N \mu$. When
$N \mu \ll 1$, the population remains essentially monomorphic, and this
is the case we focus on in this work. If instead $N \mu$ were large,
the evolutionary dynamics would occur in a polymorphic population with many 
different genotypes.

\subsection{Equilibrium sampling under stabilizing selection}
\label{subsect:sampling}

To test whether the out-of-equilibrium evolutionary dynamics
generates atypical genotypes, 
one must define the ``null hypothesis''; clearly we want to compare
to \emph{random} genotypes in 
the fitness landscape. This means that we need to sample \emph{uniformly} the
fitness landscape for each given
distance to the target phenotype; this means we focus 
on the genotypes that have a given fitness. 
The algorithmic procedure to do so is to start with any genotype in the
fitness landscape with the specified fitness and then produce a long random walk 
with importance sampling using the Metropolis Monte Carlo 
algorithm \cite{montecarlo}. In our RNA neutral ensemble context, this
simply corresponds to using the blind ant dynamics, accepting
only the mutations that do not change the distance to the target;
this is then identical to the dynamics under stabilizing selection.

From this sampling, we shall obtain equilibrium averages in this space.
We will be particularly interested in the mean mutational 
robustness, where the mutational robustness of a genotype
is defined as the fraction of the single-base mutations 
that do not change the fitness; this is the same thing as the number of neighbors 
of this genotypye that belong to the neutral ensemble divided by its total
number of neighbors.

\section{Slow evolutionary dynamics}
\label{sect:slow}
\subsection{Typical trajectories have long stasis times}

\begin{figure}[t]
 \begin{center}
 \includegraphics[width=8cm, height=11cm, angle=-90]{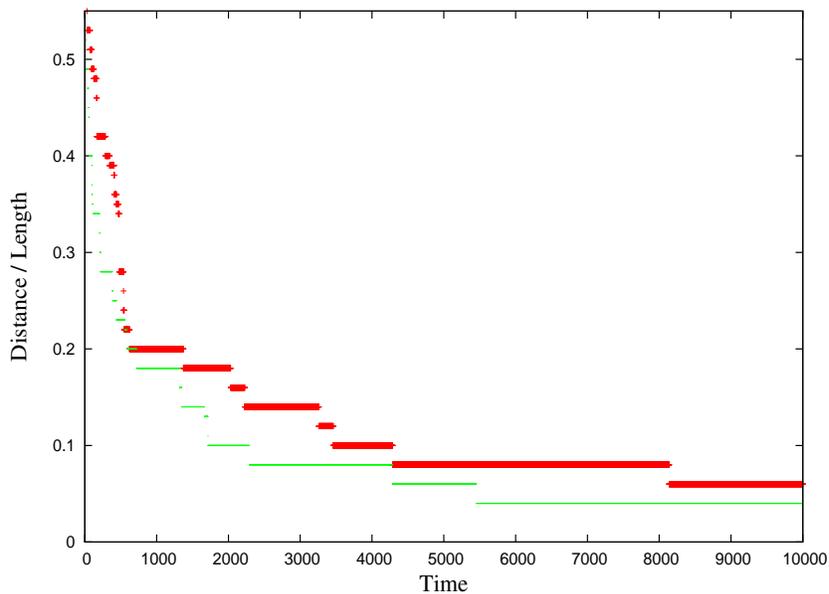}
  \end{center}
 \caption{Plot of the phenotypic distance to a target structure as a 
function of the number of mutation steps for chains of length 100.
Shown are two typical trajectories, displaying periods of stasis.
\label{fig:trajectory}
}
\end{figure}
Consider a typical evolutionary trajectory starting from a random initial
genotype. At the beginning, there is a high frequency of 
advantageous mutations, so the phenotypic distance $d_p$ to the target structure
initially decreases fast. But at long times,
as first realized in~\cite{huynenStadler1996}, the frequency
of favorable mutations becomes small and long periods
of ``stasis'' appear where the fitness remains 
constant~\cite{huynenStadler1996,fontanaSchuster1998a,fontanaSchuster1998b}.
This is illustrated for two typical evolutionary trajectories
in Fig.~\ref{fig:trajectory}.
Successive plateaus in fitness are separated by
small changes in the Hamming 
distance to the target: the steps decrease this distance 
by 1 or 2 units typically, rarely more than that.
Also, the time of a stasis period typically
increases as the distance to the target decreases: one can 
speak of diminishing returns for the effects of mutations in approaching the target.

\begin{figure}[t]
 \begin{center}
 \includegraphics[width=6cm, height=11cm, angle=-90]{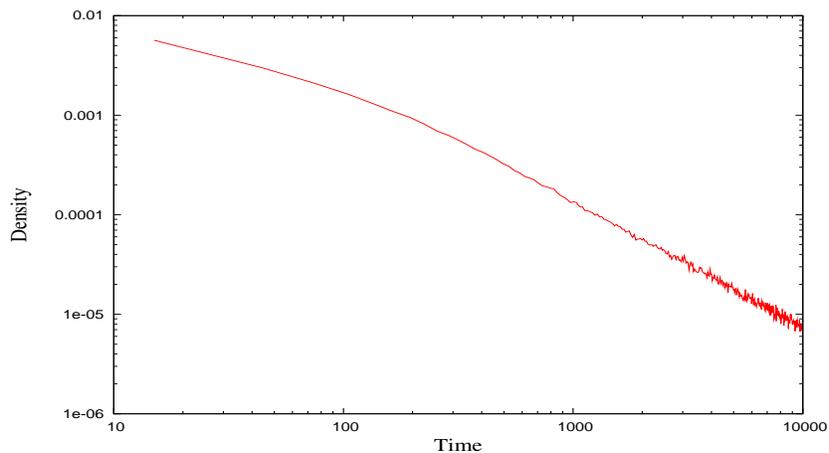}
  \end{center}
 \caption{Probability density for the times $\tau$ from the beginning of
directional selection to transition points (separating two 
periods of stasis).
   ($L=100$, averaged over 1000 trajectories for each of 30 randomly chosen targets.)
\label{fig:fat_tails}
}
\end{figure}

\subsection{Anomalously slow dynamics}

To claim that the convergence to the target is particularly slow, it is
appropriate to have a comparison benchmark. For this, consider
the situation where directional selection is for a target
genotype instead of a target phenotype. Just as for phenotypic
distances, we define the genotypic distance
of two genotypes $G_1$ and $G_2$ as $d_G = d_H(G_1,G_2)/L$
where $d_H(G_1,G_2)$ is the Hamming distance bewteen the two 
strings defining the sequences for $G_1$ and $G_2$.
If $d_G$ is then the distance between the \emph{current}
genotype and the target genotype, a mutation has
a probability $d_G/3$ to produce a strictly better genotype.
It follows that the convergence to the target is exponential
in time:
\begin{equation}
\langle d_G(t) \rangle \approx d_G(0) e^{- t/3}  \, .
\label{eq:dG_exp}
\end{equation}

Consider now the times $\tau$ when a favorable mutation arises;
a transition between one stasis period and the next generates
such an event which diminishes $d_G$ by $1/L$. If $\rho(\tau)$
is the density of these times, we have
\begin{equation}
-\frac{1}{L} \rho(\tau) = \langle d_G(\tau) - d_G(\tau-1) \rangle \approx 
\frac{d}{d \tau} \langle d_G(\tau) \rangle \, .
\end{equation}
We then obtain $\rho(\tau) = 3 L d_G(0) \exp{(-\tau/3)}$.
This shows that selection for a target genotype leads to
fast (exponential) relaxation. 

The situation is completely different
in our system where selection is instead for a target phenotype.
Guided by the above analysis, we have determined the
corresponding distribution $\rho(\tau)$ in this case. 
Fig.~\ref{fig:fat_tails} shows the result when averaging over
target and initial genotypes at $L=100$.
This distribution has a very clear fat tail which seems
compatible with a power law. This same behavior is observed
for the other values of $L$ tested (data not shown).
One can thus say that convergence to the target under
phenotypic selection is ``slow'', much slower than when compared
to genotypic selection.

\begin{figure}[t]
 \begin{center}
 \includegraphics[width=11cm, height=8cm, angle=0]{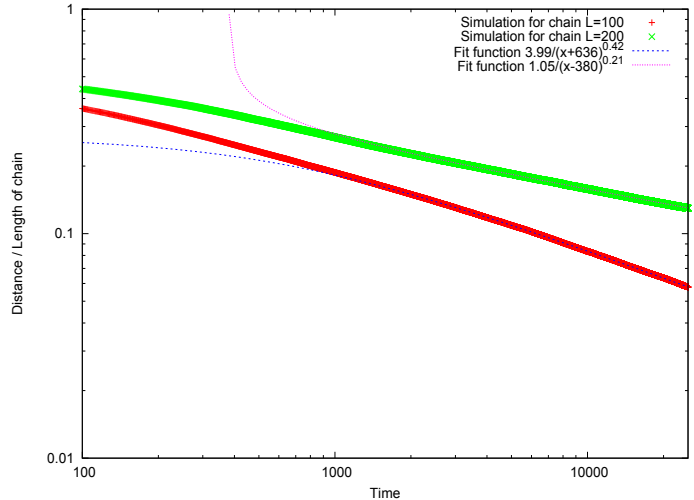}
  \end{center}
 \caption{Log-log plot of mean phenotypic distance to target structure
as a function of the number of mutational steps for chains of length 100 and 200,
when averaged over many initial genotypes and target structures. Also shown
are the fits to a shifted power law (see text).
\label{fig:logLogAverageTrajectory}
}
\end{figure}

This difference between the selection types can also be
seen by looking at the average distance to the target as
a function of evolutionary time. 
In landscapes associated with disordered systems such as glasses or spin glasses, 
the relaxation processes encountered typically have
non-exponential dynamics~\cite{young}. Empirically, two families of functions
have been used to perform fits: the stretched exponential family
which for our purposes corresponds to 
$\langle d_P \rangle \approx A \exp \left[-(t/T)^{\beta} \right]$,
and the shifted power law family for which
\begin{equation}
\langle d_P(t) \rangle \approx \frac{A}{(B + t)^{\nu}} \, .
\end{equation}
For small $L$, both types of fits lead to satisfactory
results, but at larger $L$, 
the data favor a shifted power law behavior.
In Fig.~\ref{fig:logLogAverageTrajectory}, we show that
$\langle d_P(t) \rangle$ decreases rather slowly, roughly
as an inverse power of time. This leads to a nearly straight
line on a log-log plot; also shown in that figure
are the fits to shifted power laws. This overall behavior
should be contrasted with the law 
(cf. Eq.~\ref{eq:dG_exp}) found when using selection for
a target genotype: there the approach to the target was
much faster.

\subsection{The slow dynamics is sensitive to the target}
\label{subsect:nonSelfAveraging}

\begin{figure}[t]
  \begin{center}
  \includegraphics[width=6cm, height=9cm, angle=-90]{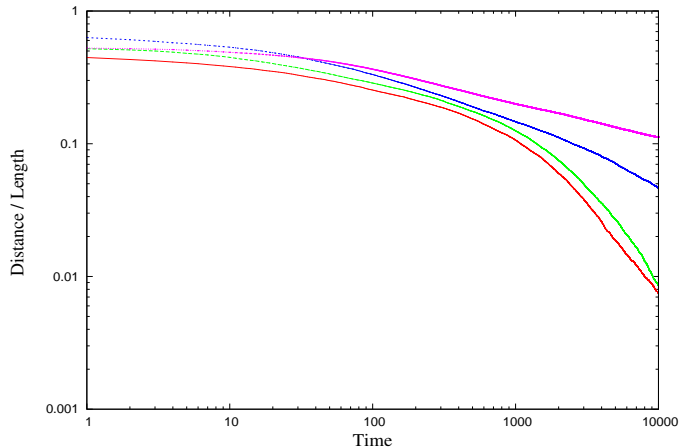}
  \end{center}
  \caption{Log-log plot of relative phenotypic distance from a target
    as a function of time for 
    simulations with different targets (all of them of
    length 60). For each target, the plotted curve is an average over
    1000 different trajectories.
 \label{fig:diff60tar}
}
 \end{figure}

To further analyse the slow approach towards a
target structure, we investigated how the evolutionary
dynamics changes with the target.
First, we ran simulations for genotypes of length 100 for
different targets. In these simulations, for each target, we
averaged the relaxation curves for different randomly
generated initial genotypes. 
Fig.~\ref{fig:diff60tar} illustrates how the relaxation
is different for different target structures.
We conclude that there is slow dynamics whose speed
depends on the target structure, a conclusion that holds for
all the values of the chain length $L$ we have 
investigated. 

Although we cannot deal with arbitrary $L$ because
of computational limitations, it is nevertheless relevant to ask
whether this dependence on targets survives for arbitrarily large $L$.
It seems possible that the relaxation behavior
is not self-averaging, \emph{i.e.}, that fluctuations associated with
different targets do not become negligible
when $L$ grows. To test this, we
carried out simulations with multiple targets for different lengths of RNA
chains, to determine whether the 
relaxation curves have smaller dispersion for the different
targets when the chain length $L$ increases. We carried out simulations for thirty
different targets, and averaged the relaxation curve for each target 
over 1000 evolutionary trajectories
with random initial genotypes. We then measured the standard deviation
$\sigma$ of the relative
distance to the target phenotype at times where the \emph{mean} distance 
(averaged over all
curves) to the target was $10$ percent of the chain length. Data are summarised
in the following table.
\begin{center}
\begin{tabular}{|l||l|l|} \hline
Chain  length & $T_{0.1}$ & $\sigma$ \\ \hline
40 & 499 & 0.0346  \\ \hline
60 & 1439 & 0.02306  \\ \hline
80 & 3349 & 0.02475 \\  \hline
100 & 6933 & 0.02683 \\   \hline
120 & 8953 & 0.02626 \\ \hline
\end{tabular}
\end{center}
Here $T_{0.1}$ denotes the mean time at which the average distance to the target
phenotype reaches the value $0.1 L$, $L$ being the chain length.
The simulations were quite time-consuming which prevented us from testing
lengths larger than 120. However, from the table we 
see that the dispersion $\sigma$ decreases
initially, but then remains practically unchanged. The relaxation process
towards different targets therefore is compatible
with a non self-averaging behavior.

\section{Weak traces of evolutionary innovations}
\label{sect:innovation}

\begin{figure}[ht]
\centering
\subfigure[Fraction of beneficial mutations.]{
 \includegraphics[width=5.5cm, height=5cm, angle=0]{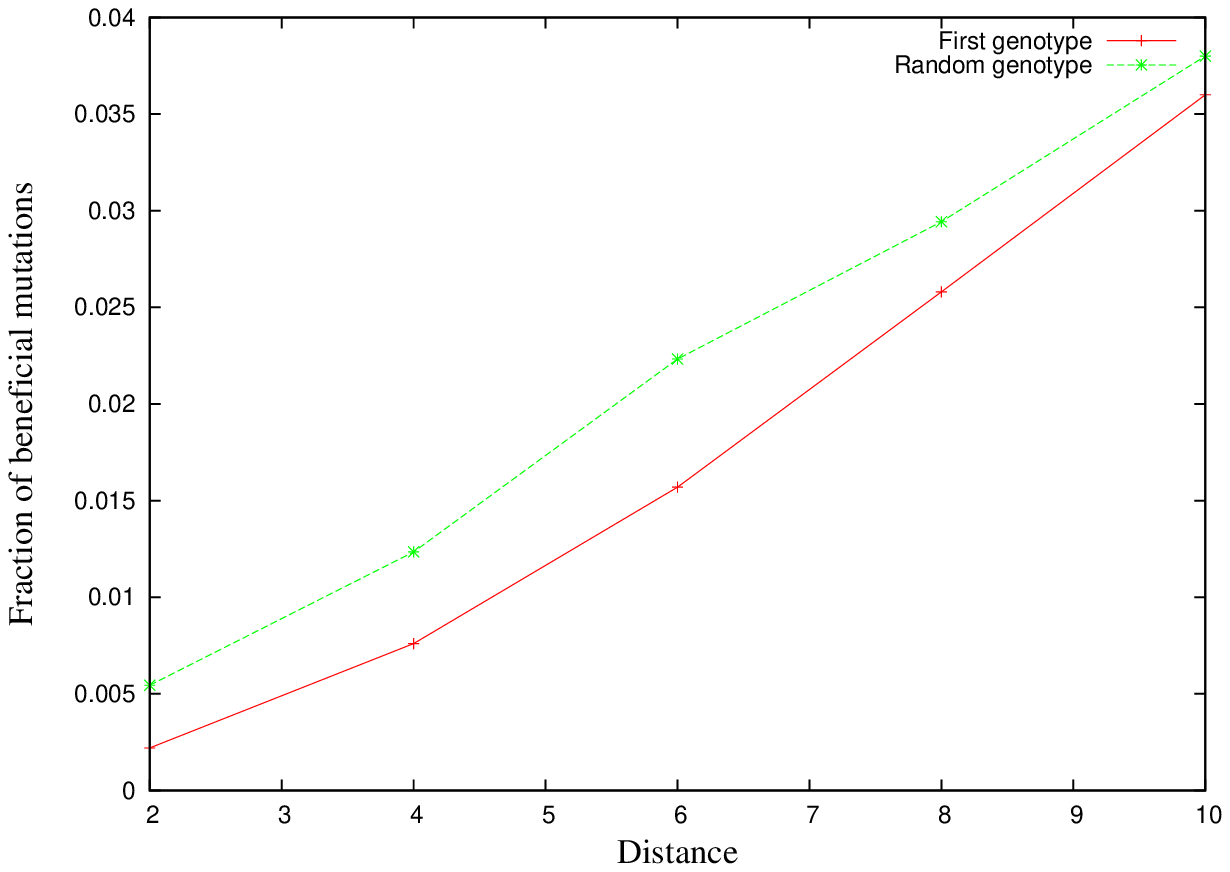}
 \label{fig:rmu_beneficial}
}
\subfigure[Fraction of deleterious mutations.]
{  
 \includegraphics[width=5.5cm, height=5cm, angle=0]{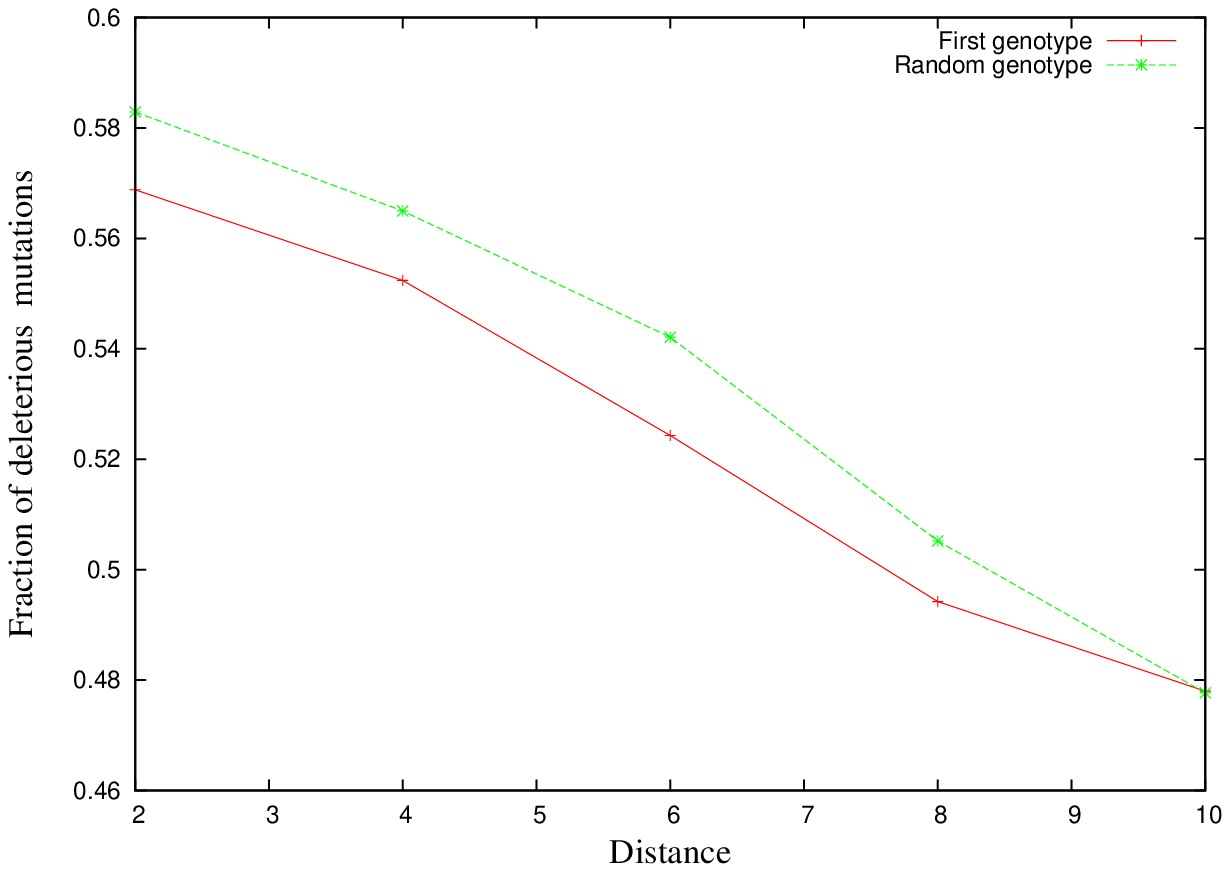}
 \label{fig:rmu_deleterious}
}
\label{fig:rmu_two}
\caption{Comparison at L=40 between entry (``first'') genotypes and random
(equilibrium) genotypes of the same fitness for the fraction of mutations that
are beneficial (a) or deleterious (b). The $x$ axis gives the phenotypic
distance to the target.}
\end{figure}

The transition from one period of stasis to the next is due
to an evolutionary ``innovation''. It is appropriate to ask
whether the associated transition genotypes can be considered
to be atypical according to some measurable quantity. Here we shall
study the \emph{short term evolvability} of
the entry genotype in a period of stasis and compare it
to that of \emph{random}
genotypes at the same distance to the target. 
For a given genotype we define its short-term 
evolvability~\cite{huyen,kirschner,meyers,cowperthwaite,wagner} as
the fraction of single base mutations that lead to a 
phenotype with higher fitness. We found that the mean fraction of such beneficial
mutations is approximately 40\% \emph{lower} for the entry genotype than 
for genotypes randomly sampled from the neutral ensemble. Thus with
our definition, the entry genotype has an atypically low evolvability.
Since it is close to a neutral ensemble of less fit phenotype,
this result is expected at a qualitative level.

However, we also find that the entry genotype has
a smaller fraction of deleterious mutations than the
neutral ensemble average. (These
mutations produce a phenotype with increased distance to the target.)
This result should be contrasted with what is naively
expected. Indeed, by construction the entry genotype has one
particular mutation which is known to be deleterious (taking
it back to the previous plateau). Neglecting all other
effects, one would predict that on average the entry
genotype would have its fraction of deleterious mutations
be $1/(3L)$ above that of genotypes in the
neutral ensemble. Instead, the effect is four times
larger and in the opposite direction.

\begin{figure}[t]
 \begin{center}
 \includegraphics[width=6cm, height=9cm, angle=-90]{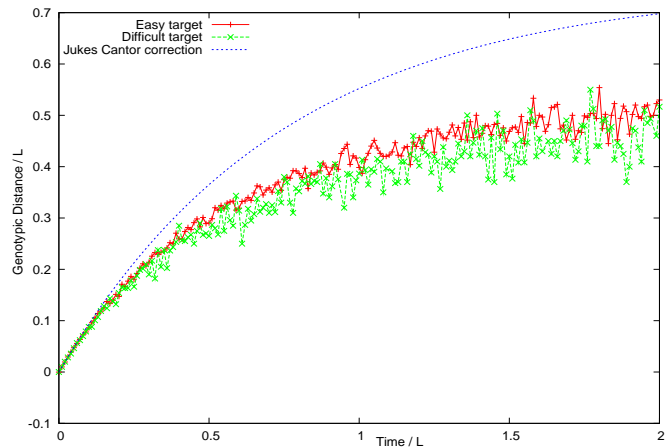}
 \end{center}
 \caption{Mean normalised genotypic distance to the
genotype arising at the beginning of the current stasis period
as a function of the number of accepted mutations divided by $L$, for chains of length $L=100$
(shown for two different targets). Also shown is the Jukes-Cantor 
correction for multiple substitutions, which is the function 
$f(x) = \frac{3}{4} \left(1 - e^{-\frac{4}{3}x}\right)$ 
used in evolutionary biology to estimate the normalised distance from the initial genotype after $x$
mutations have taken place. The function $f(x)$ approximates the amount 
of change to be expected if evolutionary change is unconstrained. Our 
curves lie somewhat below $f(x)$ which is 
expected because genotype diffusion is constrained to remain on the neutral ensemble;
nevertheless, the data show that diffusion in genotype space is rapid.
\label{fig:meangeno}
}
\end{figure}

During evolution on a plateau, one goes from the entry genotype (which we just
saw is atypical according to some objective measure)
to more random genotypes: to some extent, one looses the memory of the 
entry genotype through successive mutations. 
Is it a slow change of genotypes that is responsible for the stasis periods?
To address this question, we monitored the distance between a 
mutating genotype and the entry
genotype of the corresponding period of stasis. Because $L$ is moderately large in our simulations,
the initial growth in distance is linear in the number
of accepted mutations.  At larger times, we see a saturation effect 
caused by multiple substitutions,
as displayed in Fig.~\ref{fig:meangeno}. Overall, the evolutionary dynamics 
on the neutral ensemble shows that even in the absence of phenotypic change, 
genotypes diffuse on a neutral ensemble at a rate not much slower than 
if their diffusion was not constrained by this set, until 
they discover a new phenotype closer to the target.

To understand the expected dynamics in the absence of selection consider that 
a given site mutates with probability $1/L$ at each step, so the relaxation 
time scales as $L$. At long times one approaches the average 
distance $3/4L$. This is the behavior, translated mathematically 
in the Jukes-Cantor formula~\cite{jukes}, that the uppermost curve in Fig.\ref{fig:meangeno} represents.

In summary, 
because of the intricate relation between genotype and phenotype,
there can be slow dynamics in the approach to the target 
phenotype despite rapid evolutionary change of genotypes.
In fact, as displayed in Fig.~\ref{fig:meangeno},
the rate of change of genotypes does not seem to be significantly
different when comparing easy and difficult targets for 
the directional selection.

\section{Discussion and conclusions}

Mappings from genotypes to phenotypes play a central role in
biology, from the molecular scale up to whole organisms.
Working at the level of RNA allowed us to
use a framework for such mappings that is not biologically 
arbitrary, even though it is clearly idealised.
One of this mapping's main advantages is that it is computationally tractable. 
Within this mapping, we showed a rich phenomenology
of the evolutionary dynamics towards
an optimum phenotype: (1) the ``relaxation'' towards the target undergoes severe
slowing down as the target is approached; (2) this slowing
down gives rise to stasis periods with fat tails,
typical of what is expected in complex fitness landscapes;
(3) the relaxation curves remain sensitive to the choice of the target, even in 
the limit of long RNA sequences;
(4) the diffusion in genotype space during the periods of
stasis is not slow, in obvious contrast with what happens at the
level of phenotypes.

We observed that the probability of generating a
favorable mutation decreases severely as one approaches
the optimum, a property of the fitness landscape itself, \emph{i.e.}, the
fraction beneficial mutations goes down in this limit, much faster
than in systems undergoing exponential relaxation.
Why this leads to inverse power laws remains open,
just as in many other landscape problems coming from other fields.
Furthermore, because the stasis periods are long, 
the few innovative genotypes appearing in an evolutionary
trajectory represent a tiny fraction of the whole, and
thus most genotypes visited have little trace of 
the out-of-equilibrium dynamics. 

\bigskip

\bibliographystyle{apsrev}
%\bibliography{references}

\end{document}